\documentclass[11pt]{iopart}
\usepackage{iopams}  
\usepackage[dvips]{graphicx}
\usepackage{here}  
\begin{document}

\title{Energy loss of a heavy quark produced in a finite-size quark-gluon 
plasma}

\author{Pol Bernard Gossiaux\dag
\footnote[3]{To whom correspondence should be addressed 
(pol.gossiaux@subatech.in2p3.fr)},
J\"org Aichelin\dag, Christina Brandt\dag, Thierry Gousset\dag\ and St\'ephane 
Peign\'e\dag}

\address{\dag\ SUBATECH, UMR 6457, \'Ecole des Mines de Nantes, IN2P3/CNRS, 
Universit\'e de Nantes. 4 rue Alfred Kastler, 44307 Nantes Cedex 3, France}

\begin{abstract}
We study the energy loss of an energetic heavy quark produced 
in a high temperature quark-gluon plasma and travelling a finite distance 
before emerging in the vacuum. While the retardation time of purely collisional
energy loss is found to be of the order of the Debye screening length, we find 
that the contributions from transition radiation and the Ter-Mikayelian effect 
do not compensate, leading to a 
energy loss. reduction of the zeroth order (in an opacity expansion) energy 
loss. 
\end{abstract}

\section{Introduction}
Collisional energy loss has recently attracted some attention \cite{dars} and might be 
an explanation \cite{whdg} for the single electron puzzle observed in 
ultrarelativistic heavy ion collisions \cite{puzzle}. When incorporating collisional loss 
into quenching scenarii, one usually assumes that a heavy quark produced in a QGP
immediately undergoes stationary energy loss. In \cite{PGG}, it was however 
argued that the reaction of the medium on the heavy quark (of mass $M$ and energy $E \gg M$) can only set in 
after a retardation time $t_{\rm ret} \sim \gamma/m_D$, where $m_D$ is the Debye mass and 
$\gamma = E/M \gg 1$. A calculation 
performed in a semi-classical framework based on the collective response of the 
medium (previously used in Ref.~\cite{TG} to study the stationary collisional loss 
$-\Delta E_\infty$), indeed showed a large reduction (scaling as $\gamma$) of $-\Delta E_\infty$. 
While half of the effect could be attributed to a modification of 
initial bremsstrahlung in medium as compared to vacuum, 
the other half was interpreted as a retardation of purely collisional 
loss. In the present work we show that this latter interpretation is incorrect, 
and we instead have $t_{\rm ret} \sim 1/m_D$ for purely collisional loss, 
in agreement with the diagrammatic approach of Ref.~\cite{DG3}. We also recall the study of Ref.~\cite{GP06}, which 
does not suffer from the misinterpretation of \cite{PGG}, and presents for the first time a consistent 
calculation of heavy quark energy loss at zeroth order in an opacity expansion\footnote{Gluon radiation induced by 
quark {\it rescattering} in the QGP is not considered here, meaning that we focus on the zeroth order 
of the energy loss in an opacity expansion \cite{glv}.}, {\it i.e.} including initial 
bremsstrahlung, transition radiation, as well as purely collisional processes. 

\section{Theoretical framework and critical discussion}
We assume the heavy quark to be produced in a static QGP of high temperature $T$. The latter hypothesis 
implies the hierarchy $1/T \ll r_D \ll \lambda$, where $1/T$ is 
the average distance between two constituents of the QGP, 
$r_D=1/m_D \sim 1/(gT)$ is the Debye radius and 
$\lambda \sim 1/(g^2 T)$ is the mean free path of the heavy quark. Under these 
hypotheses, we can describe the QGP via 
its {\em collective} response to the current \cite{TG}, where the dielectric 
functions are obtained from the gluon polarization tensor. 

Let us first consider the simple case of a heavy quark produced in the past
(thus associated to a stationary classical current) in an infinite plasma.
We first solve Maxwell's equations in Fourier space and then evaluate the work 
$W$ done on the charge by the {\em induced} electric field, 
{\it i.e.} the field generated by the polarization of the medium. Identifying the quark energy loss travelling the distance $L$ in the QGP 
as $-\Delta E_{\infty}(L) = -W$ we have:
\begin{equation}
-\frac{\Delta E_{\infty}(L)}{C_F \alpha_s}=\frac{L}{v}
\int\frac{d^3 \vec{k}}{4\pi^2}\int_{-\infty}^{+\infty}
\frac{d\omega}{\omega} \left[k^2 \rho_L v_L^2+
\omega^2 \rho_T\,v_T^2\right]_{\rm ind}\,
\delta(\omega-\vec{k}\cdot\vec{v})
\label{eloss_stationn}
\end{equation}
where ${v}_L$ (${v}_T$) is the component of the quark velocity $\vec{v}$ parallel (orthogonal) to $\vec{k}$, and $\rho_{s}(\omega, k)$
(with $s=L$ or $T$) is the gluon spectral density
\begin{equation}
\hskip -2cm 
\rho_{s}(\omega, k) \equiv 2  {\rm Im}{\Delta_{s}(\omega + i\eta, k)} = 2\pi {\, \rm sgn}(\omega) \,z_s(k) 
\delta(\omega^2 -\omega_s^2(k)) + \beta_{s}(\omega, k) \theta(k^2-\omega^2) \ \ \ \  \nonumber \\
\label{spectral}
\end{equation}
where the first term corresponds to the (time-like) pole of the thermal gluon propagator $\Delta_s$, 
while the second is the magnitude 
of its cut in the space-like region and is the only term contributing to $-\Delta E_{\infty}$. 

We now turn to the case of a heavy quark produced {\it at initial time $t_{\rm prod} \sim 1/E$} 
in a hard subprocess (still in an infinite QGP). When $E$ is much larger than all other scales
(in particular $1/E \ll 1/T$), the production of the bare quark factorizes from subsequent 
evolution. The {\it kinetic} energy loss of the quark travelling the distance $L$ reads:
\begin{equation} \hskip -1cm 
-\Delta E_{\rm m,med}(L) =E_{\rm kin}(t_{\rm prod})-E_{\rm kin}(t=L/v) 
=-\int_{t_{\rm prod}}^{L/v}dt \int d^3\vec{x} \, \vec{\jmath}_1\cdot \vec{E} \ \ ,
\label{de_mech_basics}
\end{equation}
where the subscript ``m'' denotes the {\it mechanical} work done on the quark. In fact, this work already differs 
from zero for a quark produced in vacuum since initial bremsstrahlung, induced by the sudden acceleration of the quark,
leads to the reduction of $E_{\rm kin}$ after $t_{\rm prod}$. This vacuum contribution should be subtracted from 
(\ref{de_mech_basics}). When discussing jet quenching observables such as nuclear attenuation factors $R_{AA}$, 
only the vacuum-subtracted {\it medium-induced} energy loss matters. Indeed, schematically 
$R_{AA}(E) \sim d\sigma_{\rm vac}(E-[\Delta E_{\rm m,med}(\infty)- 
\Delta E_{\rm m,vac}(\infty)])/d\sigma_{\rm vac}(E)$, since the heavy quark vacuum production cross-section 
$d\sigma_{\rm vac}(E)$ already includes radiative corrections.
For the medium-induced energy loss we get:
\begin{equation} \hskip -1cm 
-\Delta E_{\rm m}\equiv -\Delta E_{\rm m,med}+ \Delta E_{\rm m,vac}
\equiv -W =-\int_{t_{\rm prod}}^{L/v}dt \int  d^3\vec{x} \, \vec{\jmath}_1\cdot \vec{E}_{\rm ind} \ \ ,
\label{de_mech_rel}
\end{equation} 
where it is again the {\em induced} electric field which enters the expression of $W$.
The calculation of (\ref{de_mech_rel}) is performed similarly to the stationary 
case\footnote{In (\ref{de_mech_basics}) and (\ref{de_mech_rel}) $\vec{\jmath}_1$ denotes the spatial component of the quark current, 
which is part of the total {\it conserved} current used in \cite{PGG}. We refer to \cite{PGG} for more details.} and we obtain \cite{PGG}:
\begin{eqnarray} 
&& \hskip -2cm -\frac{\Delta E_{\rm m}(L)}{C_F \alpha_s}= 
\int \frac{d^3\vec{k}}{2\pi^2}
\left\{ \frac{1-\cos{(kL\cos{\theta})}}{k^2+m_D^2} \right. \nonumber\\ 
&& \hskip -2cm \left. + v^2 \int_{-\infty}^{\infty} \frac{d\omega}{2\pi\omega} \left[ k^2 \cos^2{\theta} \, \rho_L+  \omega^2 \sin^2{\theta} \rho_T \right]
\times  2 \frac{\sin^2{\left((\omega-kv\cos{\theta})L/(2v)
\right)}}{(\omega-kv\cos{\theta})^2}\right\}_{\rm ind}\ \ , 
\label{mastereq2}
\end{eqnarray}
where $\theta$ is the angle between $\vec{k}$ and $\vec{v}$. Using $2\frac{\sin^2( u L/(2 v))}{u^2}
\stackrel{L\rightarrow \infty}{\approx}\frac{\pi L}{v} \delta(u)$, we find that
the dominant contribution for large $L$ is $-\Delta E_{\rm \infty}(L)$. 
Of particular phenomenological importance is the quantity $d_\infty=\lim_{L\rightarrow+\infty} 
\left(-\Delta E_{\rm m}+\Delta E_{\rm \infty}\right)$, characterizing the shift of the asymptote of 
$-\Delta E_{\rm m}$ with respect to the stationary regime $-\Delta E_{\rm \infty}$. 
A good approximation of $d_\infty$ is found to be 
$d_\infty \simeq  -C_F \alpha_s m_D \left(1+\sqrt{2}(\gamma-1)\right)$. For $\gamma\gg 1$, 
we thus obtain a fractional energy gain
$-d_\infty/E \simeq \sqrt{2} C_F \alpha_s m_D/M \approx 10-15\%$ for a charm quark 
(with $M = 1.5~{\rm GeV}$), as compared to the stationary result $-\Delta E_{\rm \infty}$. 

\vskip -0.7cm
\begin{figure}[H]
\begin{center}
\includegraphics[height=4.cm] {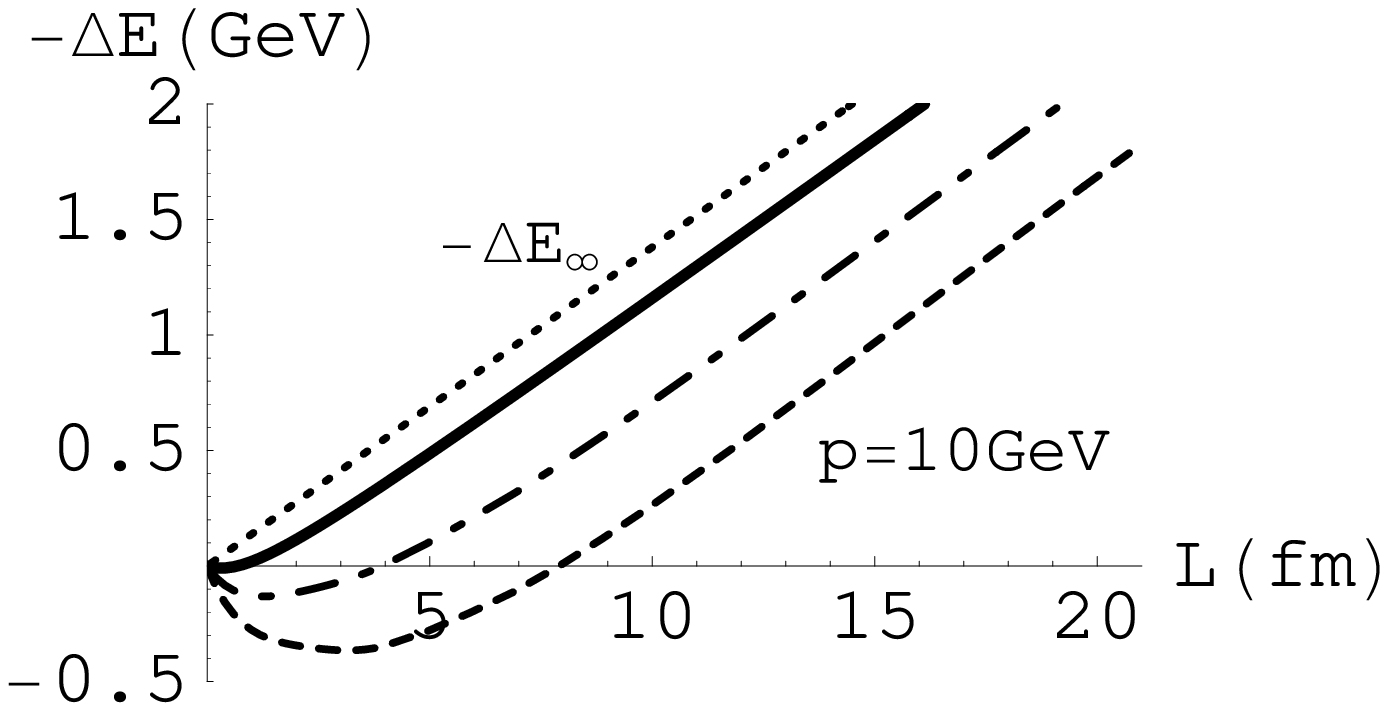} 
\includegraphics[height=4.cm] {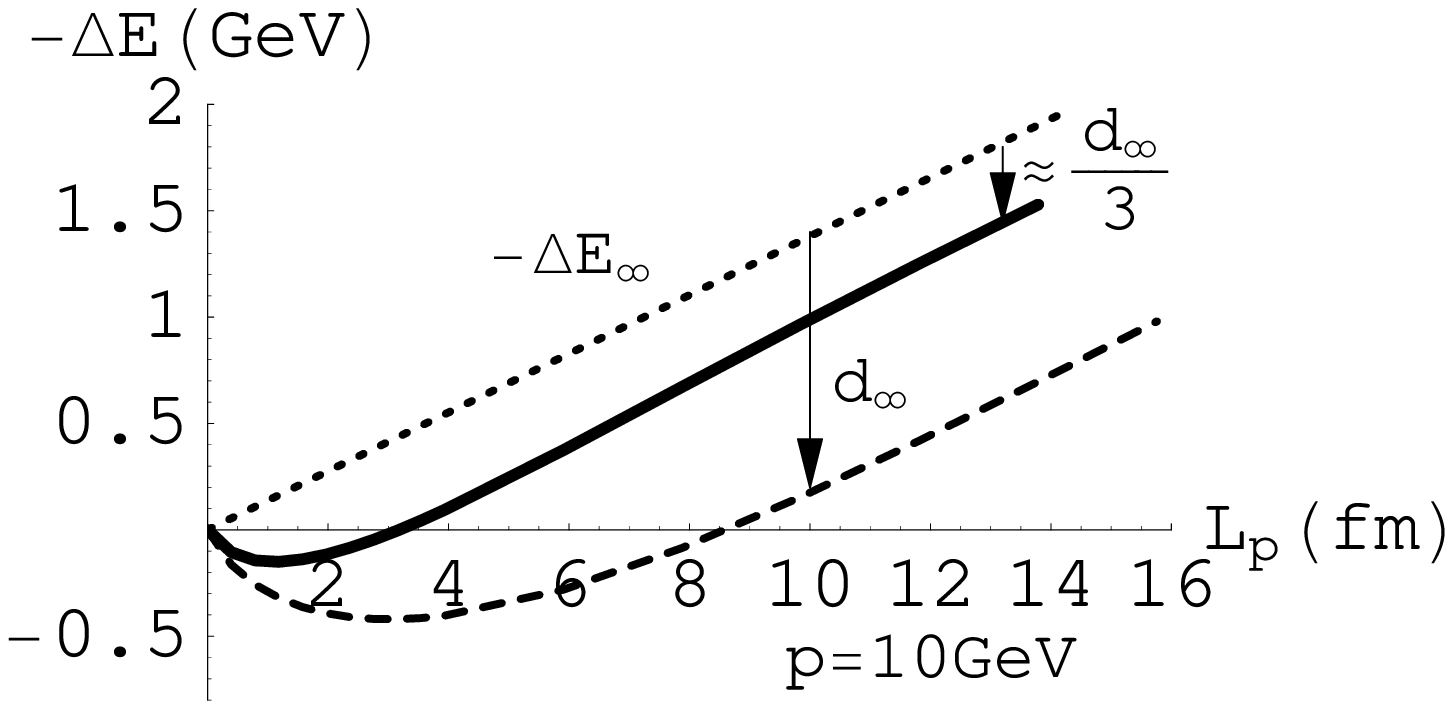} 
\end{center}
\vskip -0.7cm
\caption{\label{fig1} Energy loss as a function of the path-length in an
infinite (left) or finite (right) QGP; dotted lines ilustrate the 
stationnary collisional energy loss; dashed lines 
correspond to the mechanical work $-\Delta E_{\rm m}$ done on the charge while 
in QGP, part of it being due to the initial radiation. Left: this 
contribution is taken out on the dash-dotted line; further substracting the 
self-energy contribution, one obtains the genuine collisional energy 
loss (plain line). Right: one incorporates the mechanical work done by 
the transition field on the charge with $-\Delta E_{\rm m}$ to obtain the 
``total'' (coll. + init. rad. + trans. rad.) energy 
loss in the case of a finite-size QGP (plain line).}
\end{figure}

In Fig.~\ref{fig1} (left, dashed line), we present the energy loss (\ref{mastereq2}) for 
a charm quark of momentum $p=10~{\rm GeV}$. Although we observe a significant delay before the onset of 
stationary (linear) energy loss, this is not a genuine retardation of purely collisional loss, since various physical effects 
contribute to this delay. First, part of the work done on the charge from $t_{\rm prod}$ to $t=L/v$ 
is due to initial radiation. Due to the difference between the gluon dispersion relations in medium and in 
vacuum, the energy radiated in QGP differs from that in
vacuum. This is the QCD equivalent of the Ter-Mikayelian (TM) effect, studied in \cite{DG}
and also properly identified in \cite{PGG}. In order to single out 
contributions specific to collisional loss, we thus subtract from (\ref{mastereq2}) the TM contribution.
This is easily done by subtracting 
$\pi\,{\rm sgn}(\omega) \,z_s(k) \delta(\omega^2 -\omega_s^2(k))$ - which 
sets the gluons or plasmons on mass shell - from the spectral functions (\ref{spectral}). Denoting  
$-\Delta \tilde{E}$ and $\tilde{d}_\infty$ the quantities of interest after this subtraction, we find 
$\tilde{d}_\infty\simeq \frac{d_{\infty}}{2}$ for $\gamma\gg 1$, {\it i.e.} about half of the 
apparent ``retardation'' is due to the TM effect. In Fig.~\ref{fig1} (left), the dash-dotted
line represents $-\Delta \tilde{E}$. 

As can already be seen in the case of a charge with $v \simeq 1$ produced in 
vacuum, the radiated energy represents only $\sim$ {\em half} of the mechanical
work done on the charge after its production. Using Poynting's theorem we can 
show that  the other half is associated to the creation of the charge's proper 
field. Obviously, the {\it self-energy} contribution should not be 
counted as ``energy loss'' as it is part of the charge asymptotic state. Therefore, an accurate 
definition of energy loss - to be used for values of $L$ large enough so that 
the proper field can be disentangled from other components - is $-\Delta E(L)=-\Delta E_{\rm m}(L)-\Delta E_{\rm self}$,
where $-\Delta E_{\rm self}=E_{\rm self,vac}-E_{\rm self,med}$. Although the self-energies 
- defined as the integral of the energy density $(\epsilon E^2+\mu B^2)/2$ - 
are separately UV divergent, 
$-\Delta E_{\rm self}$ is UV convergent and $\simeq -d_\infty/2$ for $\gamma\gg 1$. 
In Fig.~\ref{fig1} (left), the plain line represents $-\Delta E(L)$ (after subtraction of the TM effect), {\it i.e.} the genuine collisional energy loss in the
case of an infinite QGP. The 
shift w.r.t. $-\Delta E_{\infty}$ appears to be $\sim \alpha_s m_D$, showing 
that the retardation time due to purely collisional processes is $t_{\rm ret} 
\sim r_D$, not $t_{\rm ret} \sim \gamma r_D$ as was argued in \cite{PGG}. 

We finally consider the most realistic situation, where the quark is produced in a hard subprocess 
in a {\it finite size} QGP and travels the distance $L_p$ before escaping the medium. 
In this case the ``asymptotic'' (we assume the quark to hadronize long after escaping the medium)
quark self-energy is the same as in vacuum, {\it i.e.} $\Delta E_{\rm self}=0$. 
It is thus legitimate to evaluate the (induced) energy loss using (\ref{de_mech_rel}), 
provided one includes the transition radiation field in the electric field $\vec{E}$. 
In Fig.~\ref{fig1} (right), the plain line presents the {\em final} result for 
the energy loss \cite{GP06}, incorporating all contributions of our model 
(collisional energy loss + initial radiation + transition radiation). 
The shift with respect to $-\Delta E_{\infty}$ 
is $\simeq d_\infty/3$, in particular it scales as $\gamma$. This is due to a 
non-compensation - already noted in \cite{DG2} -  between the TM effect and transition radiation. 
Numerically $(|d_\infty|/3)/E \simeq 3-5 \%$
(in agreement with \cite{DG2}) for $L_p> \gamma r_D$, implying an effective 
retardation time $t_{\rm ret} \simeq  4~{\rm fm}$ before the linear (stationary) regime.

\section{Conclusion}
Here we have shown that the results of \cite{PGG} (for an infinite QGP) are modified
when defining the energy loss in terms of proper asymptotic states. The retardation time of 
purely collisional loss turns out to be $t_{\rm ret} \sim r_D$. 
However, the results of \cite{GP06} for a {\it finite} QGP do not suffer from any misinterpretation
and suggest a quite large retardation time $t_{\rm ret} \sim \gamma r_D$ of the ``zeroth-order'' energy loss, including 
purely collisional processes, initial bremsstrahlung and transition radiation.

\end{document}